\documentclass[journal=jpclcd,manuscript=letter]{achemso}
\setkeys{acs}{articletitle=true}

\usepackage{amsmath}
\usepackage{amsfonts}
\usepackage{natbib}
\usepackage{amssymb}
 \usepackage[version=4]{mhchem}
\usepackage{graphicx}%
\usepackage{ulem}
\usepackage{tabularx}
\usepackage{threeparttable}
\usepackage{xcolor}
\usepackage{braket}
\usepackage{subcaption} 

\providecommand{\U}[1]{\protect\rule{.1in}{.1in}}

\definecolor{olivedrab}{rgb}{0.42, 0.56, 0.14}

\newcolumntype{F}[1]{>{\centering\arraybackslash}p{#1}}
\newcolumntype{G}[1]{>{\raggedright\arraybackslash}p{#1}}

\title{Duality of particle-hole and particle-particle theories for strongly correlated electronic systems}

\author{Aleksandra Tucholska}
\affiliation{Institute of Physics, Lodz University of Technology, ul. Wolczanska 217/221, 93-005 Lodz, Poland}

\author{Yang Guo}
\affiliation{Qingdao Institute for Theoretical and Computational Sciences, Institute of Frontier Chemistry, School of Chemistry and Chemical Engineering, Shandong University, Qingdao, Shandong 266237, China}

\author{Katarzyna Pernal}
\affiliation{Institute of Physics, Lodz University of Technology, ul. Wolczanska 217/221, 93-005 Lodz, Poland}
\email{pernalk@gmail.com}

\date{\today}

\begin{tocentry}
    \includegraphics[width=\textwidth]{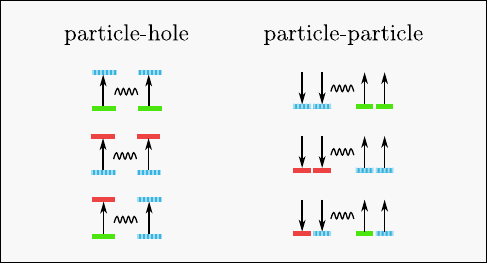}
  \end{tocentry}
\begin{document}

\begin{abstract}
We propose a novel approach to electron correlation for multireference systems. It is based on particle-hole (ph) and particle-particle (pp)
theories in the second-order, developed in the random phase approximation
(RPA) framework for multireference wavefunctions. We show a formal
correspondence 
(duality), between contributions to the correlation
energy in the ph and pp pictures. It allows us to describe correlation energy
by rigorously combining pp and ph terms, avoiding correlation double counting. 
The multireference ph,
pp, and the combined correlation methods are applied to ground and excited states
of systems in the intermediate and strong correlation regimes and compared
with the multireference second-order perturbation method (MRPT2). 
It is shown that the pp approximation fails to describe dissociation of multiple bonds.
The ph-pp combined method is overall superior to both ph and pp alone. It
parallels good accuracy of the second-order perturbation theory for ground states and singlet
excitation energies. For the singlet-triplet gaps of biradicals its accuracy
is significantly better. This is impressive, taking into account that it relies only on one- and two-body density matrices,
while MRPT2 methods typically require density matrices up to the four-body.
\end{abstract}

\maketitle
\textit{Introduction.--}
The description of electron correlation is one of the central problems of
many-body quantum theories. Accounting for electron correlation energy is
crucial for accurate predictions of electronic structure
of molecules and solids \cite{lowdin1962Review}. Particle-hole (ph) random phase approximation (RPA)
has been one of the first approaches in many-body physics to compute electron
correlation energy \cite{pines} and it has been successfully applied not only to the homogeneous electron gas \cite{gell-mann} but to real materials, see Refs.\ in \cite{rpa_rinke}. The particle-hole approximation has been vastly adopted in the framework of density functional
theory (DFT)
\cite{rpa_eshuis, rpa_rev_furche, chen2017random, julien_JCTC}. More recently a particle-particle (pp) formulation
has been developed and applied for molecules in a series of pioneering works of Yang and
coworkers \cite{van2013exchange, yang2013double, yang2013benchmark, yang2014excitation}. 
Due to its single-reference character, RPA is applicable essentially 
only to ground states and is not reliable for strongly correlated systems.
This is a serious limitation for predictions in photovoltaics, spintronics, 
and materials in excited states, as well as in systems featuring strong correlations.
For the latter case, electronic structure models are typically based on multireference (MR) models. 
Recently, efficient techniques have been developed in the framework of density matrix renormalization 
group (DMRG) theory to represent MR wavefunctions in a compact form \cite{reiherDMRG,cheng2022post}.
However, MR models miss part of the correlation energy (dynamic correlation) and various approaches have been proposed to compute it \cite{roca2012multiconfiguration}.
Among them, multireference second-order perturbation theory methods (MRPT2) \cite{GMP2,JPC94,JCP96,nevpt2,Angeli2001,rosta,SDSPT2}, remain one of the most widely used.
Depending on the formulation, they may suffer from intruder state
problem or lack of size-consistency \cite{roos1995multiconfigurational,havenith2004calibration,sen2015aspects}. 
The computational and storage costs of most MRPT2 
implementations grow rapidly with the number of 
correlated electrons and orbitals as a result of dependence on up to four-electron reduced density matrices (RDMs). 
Efforts to reduce the computational burden by
reconstructing higher-electron RDMs by 1- and 2-RDMs have not been successful to date \cite{zgid2009study}. Multi-reference methods that require only 1- to 3-RDMs have also been reported by Chan,\cite{icMRCC-CT} Evangelista,\cite{DSRGrev2014} and Yanai\cite{kurashige2014complete}. 
Recently a multireference particle-hole approximation for electron correlation has been developed in the
framework of the adiabatic connection (AC) theory \cite{ac_prl,Pernal:18b,Pastorczak:18a}, which only needs 1- and 2-RDMs. The AC method has been
successfully applied with complete active space (CAS) and DMRG to
describe ground states and multiplet splittings \cite{pastorczak2019capturing,Beran:21}. The method is less
accurate for excited states, leading to overestimation of the excitation
energies. A correction proposed by some of us partially amends this deficiency but it requires multiple calculations \cite{drwal2021}. Until now, a multireference particle-particle approximation has not been known.

The purpose of this work is to develop a unified framework for multireference particle-hole and particle-particle approximations. We propose a
rigorous ph-pp combined method that accurately predicts correlation energy in ground and excited state systems based solely on 1- and 2-RDMs. 
This is achieved by deriving ph and pp
correlation energy formulae which are consistent with the second-order
correlation energy and by revealing a nontrivial duality of the ph and pp
contributions.

\textit{Theory.--}
 To compute correlation energy, one typically starts with a qualitatively correct reference wavefunction, $|\Psi_{0}^{(0)}\rangle$, capturing static correlation within an $N$-electron Hilbert space. 
 The subscript "0" denotes a target $N$-electron, ground or excited, electronic state under consideration. 
 Introduce a partially-interacting Hamiltonian
\begin{equation}
\forall_{\alpha\in\lbrack0,1]}\ \ \ \hat{H}^{\alpha}=\hat{H}^{(0)}+\alpha
\hat{H}^{\prime} \  , \label{Hal}%
\end{equation}
and  let $\{\vert \Psi_{\nu}^{\alpha}\rangle \}$ represent a set of all eigenstates of $\hat{H}^{\alpha}$.
$M$-electron states form a resolution of identity (RI), $\hat{1}_M=\sum_{\nu \in {\cal{H}}_M}\vert \Psi_\nu^{\alpha}\rangle \langle \Psi_\nu^{\alpha}\vert$, ${\cal H}_M$ denotes an $M$-electron Hilbert space. The second-order correlation energy formally reads%
\begin{equation}
E_{0}^{(2)}=\left\langle \Psi_{0}^{(0)}|\hat{H}^{\prime}|\Psi_{0}
^{(1)}\right\rangle \ \ \ . \label{EPT2}%
\end{equation}
The 1st-order wavefunction, $|\Psi_{0}
^{(1)}\rangle$, typically consists of determinants or configuration state functions in the 1st-order interaction space \cite{Dyall1995,Angeli2001}.
We will show that the PT2  correlation energy can be turned into equivalent forms related to a two-fermionic (ff), either ph or pp, picture.

Assume that $\hat{H}^{\prime}$ includes one- and two-electron operators%
\begin{equation}
\hat{H}^{\prime}=\sum_{pq}{}^{\prime}\ h_{pq}\hat{a}_{p}^{\dagger}\hat{a}%
_{q}+\frac{1}{2}\sum_{pqrs}{}^{\prime}\ \left\langle pq|rs\right\rangle
\hat{a}_{p}^{\dagger}\hat{a}_{q}^{\dagger}\hat{a}_{s}\hat{a}_{r}%
\  ,\label{Hp}%
\end{equation}
where 
$pqrs$ are general spinorbital indices, $h_{pq}$ and $\left\langle pq|rs\right\rangle $ denote,
respectively, one- and two-electron integrals, assumed in this work to be
real-valued. Primes in Eq.(\ref{Hp}) indicate restrictions on the summations inferred from the adopted form of $\hat H^{(0)}$. Consider two formal relations, obtained by
using anticommutator rules for fermionic operators and the RI for $N$- and
$N+2$-electron states
\begin{equation}
\hat{a}_{p}^{\dagger}\hat{a}_{q}^{\dagger}\hat{a}_{s}\hat{a}_{r}= \sum_{\nu \in {\cal{H}}_{N}} \hat{a}_{p}^{\dagger}\hat{a}_{r}\left\vert \Psi_{\nu}^{\alpha
}\right\rangle \left\langle \Psi_{\nu}^{\alpha}\right\vert \hat{a}%
_{q}^{\dagger}\hat{a}_{s}-\hat{a}_{p}^{\dagger}\hat{a}_{s}\delta
_{qr}\ \ \ ,\label{McB}%
\end{equation}
and%
\begin{align}
\hat{a}_{p}^{\dagger}\hat{a}_{q}^{\dagger}\hat{a}_{s}\hat{a}_{r} &  = \sum
_{\nu \in {\cal{H}}_{N+2}}  \hat{a}_{r}\hat{a}_{s}\left\vert \Psi_{\nu}^{\alpha}\right\rangle
\left\langle \Psi_{\nu}^{\alpha}\right\vert \hat{a}_{q}^{\dagger}\hat{a}%
_{p}^{\dagger}+\delta_{sq}\hat{a}_{p}^{\dagger}\hat{a}_{r}
\nonumber\\
&-\delta_{ps}\hat
{a}_{q}^{\dagger}\hat{a}_{r}
  +\delta_{rq}\hat{a}_{s}\hat{a}_{p}^{\dagger}-\delta_{pr}\hat{a}_{s}\hat
{a}_{q}^{\dagger}\ \ \ .\label{McBpp}%
\end{align}
Employing Eqs.(\ref{McB}) or (\ref{McBpp})  in
Eq.(\ref{Hp}) turns Eq.(\ref{EPT2}) into the ph and pp
correlation energy expressions reading
\begin{equation}
E_{\text{corr}}^{\text{ph}}=\frac{1}{2}\sum_{pqrs}{}^{\prime}\ \left\langle
pq|rs\right\rangle  \sum
_{\nu \in {\cal{H}}_{N}}  \left[  \mathbf{\gamma}_{\text{ph}}^{\nu}\right]
_{pr}^{(0)}\left[  \mathbf{\gamma}_{\text{ph}}^{\nu}\right]  _{qs}%
^{(1)}\ \ \ \label{E2ph}%
\end{equation}
and
\begin{equation}
E_{\text{corr}}^{\text{pp}}=\frac{1}{2}\sum_{pqrs}{}^{\prime}\ \left\langle
pq|rs\right\rangle \ \sum
_{\nu \in {\cal{H}}_{N+2}}  \left[  \mathbf{\gamma}_{\text{pp}}^{\nu
}\right]  _{pq}^{(0)}\left[  \mathbf{\gamma}_{\text{pp}}^{\nu}\right]
_{rs}^{(1)}\ \ \ ,\label{E2pp}%
\end{equation}
respectively. 
The ph transition reduced density matrices (TRDMs),
$\mathbf{\gamma}_{\text{ph}}^{\nu}$, pertain to electron-number-conserving
transitions between states $0$ and $\nu$, while their pp counterparts,
$\mathbf{\gamma}_{\text{pp}}^{\nu}$, connect $0$ and $\nu$ states differing in
the number of electrons by $2$. In the $0$th- and $1$st-order, 
TRDMs are given as%
\begin{gather}
\left[  \mathbf{\gamma}_{\text{ff}}^{\nu}\right]  _{I}^{(0)}=\left\langle
\Psi_{0}^{(0)}|\hat{o}_{I}|\Psi_{\nu}^{(0)}\right\rangle \ \ \ ,\label{TRDM0}%
\\
\left[  \mathbf{\gamma}_{\text{ff}}^{\nu}\right]  _{I}^{(1)}=\left\langle
\Psi_{0}^{(0)}|\hat{o}_{I}|\Psi_{\nu}^{(1)}\right\rangle +\left\langle
\Psi_{0}^{(1)}|\hat{o}_{I}|\Psi_{\nu}^{(0)}\right\rangle \ \ \ ,\label{TRDM1}%
\\
\forall_{I=pq}\ \ \ \hat{o}_{I}=\left\{
\begin{array}
[c]{cc}%
\hat{a}_{q}^{\dagger}\hat{a}_{p} & \ \ \ \text{ff}=\text{ph}\\
\hat{a}_{p}\hat{a}_{q} & \ \ \ \text{ff}=\text{pp}%
\end{array}
\right.  \ \ \ ,\label{oI}%
\end{gather}
where 
$I$ is a compound index for
an ordered pair of spinorbital indices and the operator $\hat{o}_{I}$ is
of the ph or pp type.  In deriving Eqs.(\ref{E2ph}) and (\ref{E2pp}), it has
been assumed that a $1$st-order correction to 1-RDM vanishes. The latter is exactly satisfied for the Hartree-Fock state. In the case of multireference states, the $1$st-order correction to 1-RDM is negligible comparing to higher-order corrections \cite{matouvsek2023toward}.
Taking it into account, 
the equivalence relations follow
\begin{gather}
\left\langle \Psi_{0}^{(0)}|\hat{a}_{p}^{\dagger}\hat{a}_{q}^{\dagger}\hat
{a}_{s}\hat{a}_{r}|\Psi_{0}^{(1)}\right\rangle =\nonumber\\
 \sum
_{\nu \in {\cal{H}}_{N+2}} \left[  \mathbf{\gamma}_{\text{pp}}^{\nu}\right]  _{pq}^{(0)}\left[
\mathbf{\gamma}_{\text{pp}}^{\nu}\right]  _{rs}^{(1)}= \sum
_{\nu \in {\cal{H}}_{N}} \left[
\mathbf{\gamma}_{\text{ph}}^{\nu}\right]  _{pr}^{(0)}\left[  \mathbf{\gamma
}_{\text{ph}}^{\nu}\right]  _{qs}^{(1)} .\label{equiv}%
\end{gather}
They pave a way to combining ph and pp approaches in a rigorous manner, without correlation double counting (i.e.\ without accounting for the same effect twice in the correlation energy expression). 

One should recall that 
Eq.(\ref{E2ph}) has been already derived within the, formally exact, ph adiabatic connection formalism \cite{ac_prl} approximated at lowest order. 
Analogously, by employing the representation of a two-electron operator given in Eq.(\ref{McBpp}) in AC, one would arrive at Eq.(\ref{E2pp}) in the lowest-order AC.

We utilize a CAS ansatz \cite{Roos1987}, remaining the most widely used model to capture the static correlation of strongly correlated systems \cite{olsen2011casscf}, as $\Psi_{0}^{(0)}$. It includes doubly occupied and partially occupied orbitals, belonging to sets (o) and (a), respectively. The remaining orbitals are unoccupied (virtual) and belong to a set (v). 
Dyall's Hamiltonian \cite{Dyall1995} reading  $\hat{H}^{(0)}=$ $\sum_{pq\in {\rm (o)}}h_{pq}^{\text{eff}%
}\hat{a}_{p}^{\dagger}\hat{a}_{q}+\sum_{pq\in {\rm (v) }}h_{pq}^{\text{eff}}\hat{a}%
_{p}^{\dagger}\hat{a}_{q}+\sum_{pq\in {\rm (a)} }h_{pq}\hat{a}%
_{p}^{\dagger}\hat{a}_{q}+\frac{1}{2}\sum_{pqrs\in {\rm (a)} }\left\langle
pq|rs\right\rangle \hat{a}_{p}^{\dagger}\hat{a}_{q}^{\dagger}\hat{a}_{s}%
\hat{a}_{r}$ [for a definition of $h^{\rm eff}$ see Eq.(S.7) in the Supplemental Material]
is adopted.
It  has been used for developing perturbation theories in 
various contraction variants \cite{Dyall1995,Angeli2001}. From now on the acronym PT2 will denote a partially-contracted NEVPT2 method \cite{Angeli2001},  used in this work. 

A general framework for
developing approximations for transition properties is provided by the Rowe's
equation of motion (EOM) \cite{rowe}. In EOM written for the $\alpha
$-dependent Hamiltonian, Eq.(\ref{Hal}), an excitation
operator $\hat{O}_{\nu}^{\alpha}$ generates a $\nu$th state 
as $\hat{O}_{\nu}^{\alpha}\vert \Psi_{0}^{(0)}\rangle =\vert
\Psi_{\nu}^{\alpha}\rangle $. In the random phase approximation, $\hat{O}_{\nu}^{\alpha}$ includes products of only two fermionic (ff) operators,
namely $\hat{O}_{\nu}^{\alpha}=\sum_{I}\left[  \mathbf{Z}_{\nu}^{\alpha
}\right]  _{I}\hat{o}_{I}^{\dagger}$, where $\hat{o}_{I}^{\dagger}$ is an adjoint of the ph or pp operator defined in Eq.(\ref{oI}). The resulting EOM takes the form of a generalized eigenvalue problem%
\begin{equation}
\begin{split}
 &\sum_{J} A^{\alpha}_{IJ}\left[  \mathbf{Z}_{\nu}^{\alpha}\right]  _{J}
 =\omega_{\nu}^{\alpha}\sum_{J}
S_{IJ}  \left[  \mathbf{Z}_{\nu}^{\alpha}\right]  _{J}%
\ ,\label{REOM}%
\\&
A^{\alpha}_{IJ}=\left\langle \Psi_{0}^{(0)}\right\vert \left[  \hat{o}_{I},[\hat
{H}^{\alpha},\hat{o}_{J}^{\dagger}]\right]  \left\vert \Psi_{0}^{(0)}%
\right\rangle\ ,
\\&
S_{IJ} = \left\langle \Psi_{0}^{(0)}\right\vert \left[  \hat{o}_{I},\hat{o}_{J}^{\dagger}\right]  \left\vert \Psi_{0}^{(0)}%
\right\rangle\ ,
\end{split}
\end{equation}
where the eigenvalues $\omega_{\nu}^{\alpha}$ are transition energies,
while the
eigenvectors $\mathbf{Z}_{\nu}^{\alpha}$ multiplied by the metric matrix 
$\mathbf{S}$
yield
transition density matrices
$\left[  \gamma_{\text{ff}}^{\nu}\right]_{I}^{\alpha}=\sum_{J} S_{IJ} \left[  \mathbf{Z}_{\nu}^{\alpha}\right]
_{J}$.
Notice that the EOM has
been derived by assuming 
a killer condition, 
$[  \hat{O}_{\nu}^{\alpha}]^{\dagger}\vert \Psi_{0}^{(0)}\rangle =0$.
The EOM eigenproblem in Eq.(\ref{REOM}) is symplectic. The two sets of eigenvectors in the ph variant correspond to excited and de-excited N-electron states (the corresponding eigenvalues are of the same magnitude and opposite signs). The pp equations yield eigenvectors describing transitions to $N+2$ and $N-2$-electron states and the eigenvalues correspond to double-electron attachment/detachment energies.  

\begin{figure}[h]
    \centering
    \includegraphics[width=0.7\textwidth]{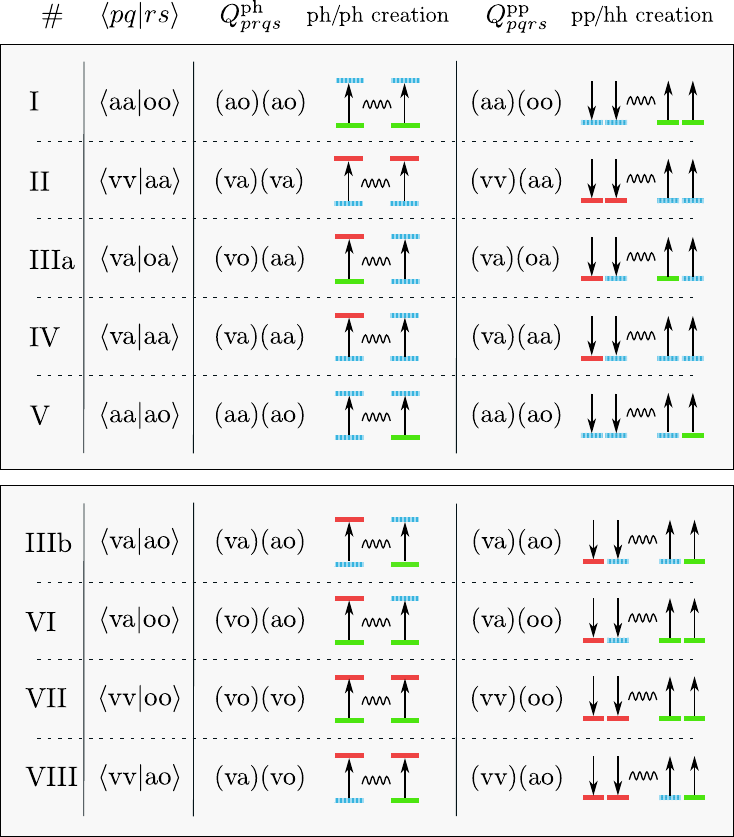}
    \caption{Upper panel: correspondence between dual  ph and pp terms giving numerically different contributions to the correlation energy.
    Lower panel: Correspondence between dual ph and pp 'EKT' terms  giving numerically equivalent contributions to correlation.
    A contribution to the correlation energy, 
     see Eq.(\ref{Eff}), 
    is given by a $Q^{\text{ff}}$ element multiplied by a corresponding two-electron integral. Single up-arrows denote excitations to N-electron states, double up-arrows and double down-arrows denote transitions to N$-2$ and N+2-electron states, respectively. }
    \label{fig:diag-phppEKT}
\end{figure}

At $\alpha=0$ the EOM main matrix is block-diagonal if CAS wavefunction and the Dyall Hamiltonian are used \cite{guo2024spinless, Pastorczak:18a}. 
Consequently, the ph and pp
correlation energy expressions, Eq.(\ref{E2ph}) and (\ref{E2pp}), can be written in a
common form as
\begin{equation}
E_{\text{corr}}^{\text{ff}}=\frac{1}{2}\sum_{I,J\in\Omega}\ \left\langle I|J\right\rangle
\ Q_{\mathcal{P}_{\text{ff}}(IJ)}^{\text{ff}}\ \ \ .
\label{Eff}
\end{equation}
$\left\langle I|J\right\rangle $ denotes a two-electron integral, i.e.
if $I=pq$ and $J=rs$ then $\left\langle I|J\right\rangle =\left\langle
pq|rs\right\rangle $, $\mathcal{P}_{\text{ff}}(IJ)$ permutes indices
encoded by $I$ and $J$ specifically to either ph or pp, namely%
\begin{equation}
\forall_{\substack{I=pq\\J=rs}}\ \ \ \mathcal{P}_{\text{ff}}(IJ)=\left\{
\begin{array}
[c]{cc}%
prqs & \ \ \ \text{ff}=\text{ph}\\
pqrs & \ \ \ \text{ff}=\text{pp}%
\end{array}
\right.  \ \ \ 
\end{equation}
and pairs of indices $I,J$ are from the set
\begin{equation}
    \Omega = \{ \text{(vv), (oo), (va), (ao), (aa)}\} \ \ \ ,
    \label{Omega}
\end{equation}
where, e.g.\ a subset (va) contains pairs of virtual and active  orbitals, 
analogous notation is used for the other subsets.
All integrals which are included in Eq.(\ref{Eff}) are listed in Fig.~\ref{fig:diag-phppEKT}.
The $\mathbf{Q}^{\text{ff}}$ matrices 
follow from the 1st-order perturbation theory applied to Eq.(\ref{REOM}) and are given by the $0$th-order TRDMs, and the coupling
matrix $\mathbf{D}^{\text{ff}}$ 
\begin{equation}
Q_{pqrs}^{\text{ff}}= \sum_i \sum_{\substack{\mu \in {\cal H}_M\\\nu \in {\cal H}_{M'}}}
\left[  \mathbf{\gamma}_{\text{ff}}^{\mu
}\right]_{p_iq_i}^{(0)} D_{\mu\nu}^{\text{ff}}\left[  \mathbf{\gamma}_{\text{ff}%
}^{\nu}\right]  _{r_is_i}^{(0)}\ \ \ ,\label{QIJ}%
\end{equation}
where $M,M'$ are equal to $N$ for ff=ph and $M=N+2$, $M'=N-2$ in the pp approach (ff=pp). The summation with respect to $i$ denotes all unique permutations $p\leftrightarrow q$, $r\leftrightarrow s$, and $pq\leftrightarrow rs$, see Supplemental Material (Section 2) for examples. 
For explicit expressions of the $Q_{pqrs}^{\text{ff}}$ in the ph approximation see Ref.\cite{Pastorczak:18a}. In the pp variant the $Q^{\text{pp}}$ elements are given as
\begin{align}
Q_{pqrs}^{\text{pp}}=&\sum_{\substack{\mu\in\mathcal{H}_{N+2}\\
\nu\in\mathcal{H}_{N-2}}}\left[  \mathbf{\gamma}_{\text{pp}}^{\mu
}\right]  _{pq}^{(0)} 
D_{\mu\nu}^{\text{pp}}
\left[  \mathbf{\gamma}_{\text{pp}%
}^{\nu}\right]  _{rs}^{(0)}\\=&
\sum_{\substack{\mu\in\mathcal{H}_{N+2}\\
\nu\in\mathcal{H}_{N-2}}}\sum_{IJ}
\frac{
[  \mathbf{Z}_{\mu}^{(0)}]_{I}
A^{(1)}_{IJ}
[  \mathbf{Z}_{\nu}^{(0)}]_{J}
}
{\omega_{\mu}^{(0)}-\omega_{\nu}^{(0)}}
[\mathbf{SZ}_{\mu}^{(0)}]_{pq}
 [\mathbf{SZ}_{\nu}^{(0)}]_{rs} \ \ \nonumber ,
\label{Qpp-omeq}
\end{align}
 $\mu\in\mathcal{H}_{N+2}$ and $\nu\in\mathcal{H}_{N-2}$ pertain to the subsets of eigenvectors $\mathbf{Z}^{(0)}_\nu$ that satisfy the respective inequalities, \cite{van2013exchange, van2014exchange}
\begin{equation}
\begin{split}
     &  (\mathbf{Z}^{(0)}_{\mu})^\dagger \mathbf{S} \mathbf{Z}^{(0)}_{\mu} > 0 \ \ \ , \\
     &  (\mathbf{Z}^{(0)}_{\nu})^\dagger \mathbf{S}\mathbf{Z}^{(0)}_{\nu} < 0 \ \ \ ,
\end{split}
\label{SI-ineq}
\end{equation}
and the eigenvalues $\omega^{(0)}_{\mu\in\mathcal{H}_{N+2}}$ and $\omega^{(0)}_{\nu\in\mathcal{H}_{N-2}}$ describe transitions from $N$- to $N+2$- and $N-2$-electron states, respectively. 
For the $\mu\in\mathcal{H}_{N+2}$ solutions, only the (vv), (va), (av) and (aa) elements of eigenvectors are nonzero, while for $\nu\in\mathcal{H}_{N-2}$ states, the nonvanishing eigenvector elements are of the (oo), (ao), (oa) and (aa) type.
$A^{(1)}$ is the first order ppEOM matrix, cf.\ Eq.(\ref{REOM}). For explicit expression for  $\alpha$-dependent $A$ matrix in terms of one- and two-electron reduced density matrices see Eq.(S2) in the Supplemental Material.

The ff=ph correlation energy expression shown in Eq.(\ref{Eff}) is the
familiar AC0 approximation, derived originally within the adiabatic connection
framework \cite{ac_prl, Pernal:18b}. The ff=pp multireference correlation energy expression is shown for the first time. For consistency with the previous works, we will denote Eq.(\ref{Eff}) as phAC0 for ff=ph and, by analogy, an ff=pp variant will be called ppAC0. 

Using the equivalence relation from Eq.(\ref{equiv}) 
one can identify a duality between elements $Q_{IJ}^{\text{ff}}$ computed either with ph or pp approximations which multiply the same two-electron integral. Dual terms are presented in Fig.~\ref{fig:diag-phppEKT}. They represent different physics in ph and pp theories.
In the ph theory, correlation arises from the coupled creation of two particle-hole pairs, ph-ph coupling, illustrated as two single upward 
arrows in the ph panel of Fig.~\ref{fig:diag-phppEKT}. Conversely, 
in the pp theory, correlation involves 
creation of two particles coupled with creation of two holes, pp-hh coupling, represented by double down- and double up-arrows in respective diagrams.
The revealed ph-pp duality allows one to combine both theories by including either ph or pp contributions out of a dual pair and obtaining a new expression for correlation energy. 

It can be shown by inspection that the (ao) and (va) blocks of 0th-order EOMs reduce to the extended Koopmans' theory (EKT) equations describing single ionization and single electron attachment processes \cite{nevpt2}, in both ph and pp variants.  
Notice that they  are also equal to their counterparts in the NEVPT2 theory \cite{Angeli2001, angeli2001n, guo2024spinless}. 
Dual ph and pp contributions to the correlation energy, presented in the lower panel of Fig.~\ref{fig:diag-phppEKT} as terms IIIb, VI, VII, VIII,  are equal (see also Table S2 in the Supplemental Material for the numerical evidence). They will be referred to as EKT terms. 

\begin{figure}[h]
    \centering
    \includegraphics[width=0.98\columnwidth]{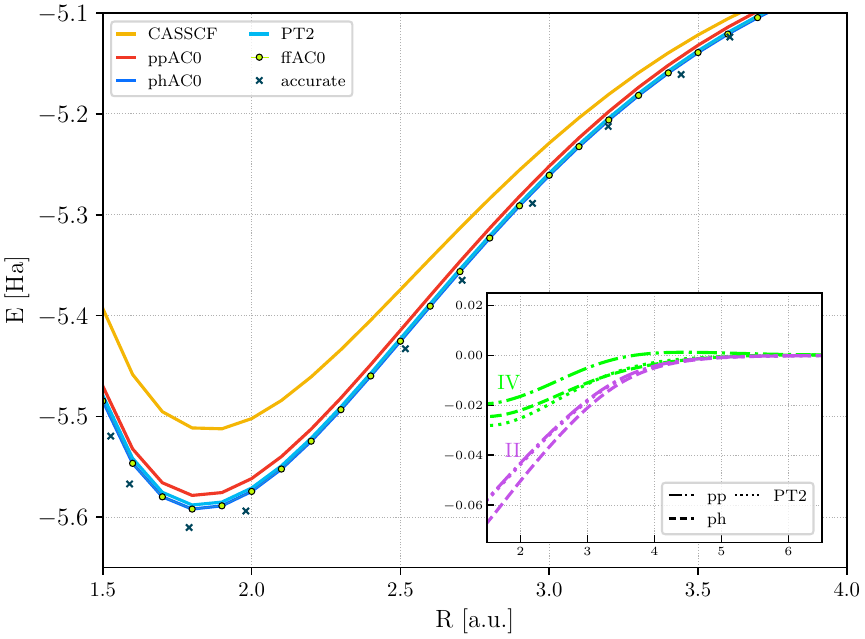}
    \caption{Total energies of $\ce{H10}$ in cc-pVDZ basis \cite{Dunning:89}, calculated with CASSCF(10,10) as a reference wavefunction. Inset panel contains contributions II and IV, see Fig.~\ref{fig:diag-phppEKT},
  calculated   for  ppAC0 (dashdotted), phAC0 (dashed), and  PT2 (dotted). Accurate results from Ref.~\cite{eriksen2019many}.}
    \label{fig:h10}
\end{figure}

\begin{figure}[h]
    \begin{center}
    \includegraphics[width=0.98\columnwidth]{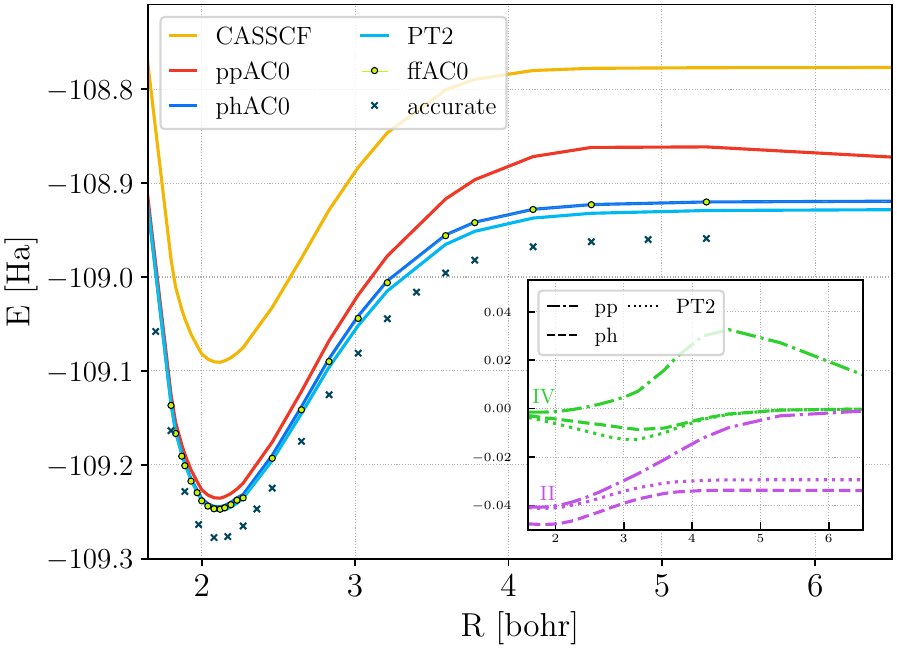}
     \caption{Total energies of $\ce{N2}$ in cc-pVDZ basis \cite{Dunning:89}, calculated with CASSCF(6,6) as a reference wavefunction. Inset panel contains contributions II and IV, see Fig.~\ref{fig:diag-phppEKT},
  calculated for  ppAC0 (dashdotted), phAC0 (dashed), and  PT2 (dotted). 
  Accurate results are taken from FCI calculations Ref.~\cite{eriksen2019many}.
   }
    \label{fig:n2}
     \end{center}
\end{figure}

Dual terms yielding numerically different contributions to the correlation energy in the ph and pp variants, are shown in the upper panel of Fig.~\ref{fig:diag-phppEKT}.
The anticipated superiority of either ph or pp approximations for computing
contributions I, II, and IIIa can be inferred from theoretical arguments.
Recall that the killer condition is assumed in EOMs. Considering the pp equations, this condition is satisfied  in the $0$th-order
if the 
$N+2$- or $N-2$-electron states 
do not belong to the active
sector, $\nu\notin$(aa), i.e.\ they do not result from attaching/detaching two
electrons to/from the active space:%
\begin{equation}
\nu\notin\text{(aa)}\wedge\nu\in\mathcal{H}_{N+2}\cup\mathcal{H}%
_{N-2}\Longrightarrow\left[  \hat{O}_{\nu}^{(0)}\right]  ^{\dagger}\left\vert
\Psi_{0}^{(0)}\right\rangle =0\ \ \ .
\label{killer}
\end{equation}
For
a ph excitation operator
the killer condition is violated but a weaker condition -
orthogonality of states, $\langle \Psi_{0}^{(0)}\vert\hat{O}_{\nu}^{(0)}\vert \Psi_{0}%
^{(0)}\rangle =0$, 
is fulfilled unless $\hat{O}_{\nu}^{(0)}$  includes $\hat{o}_{pq}$ excitations where the indices $pq$ are from the (aa) space.
The ph approach is therefore expected to be more accurate than pp for contributions I and II, 
since in the latter picture the pertinent terms involve creation of two fermions in the (aa) space.
The pp approach is favored over ph for IIIa terms since the killer condition is satisfied in pp-EOM for the (va) and (oa) excitations, see Eq.(\ref{killer}).
Concerning the ph-pp dual terms
with three active indices, terms IV and V, numerical analysis obtained for the paradigmatic strongly correlated molecules:\ N$_{2}$ dimer and H$_{10}$ chain with symmetrically elongated bonds,
see Figs.~\ref{fig:h10} and \ref{fig:n2}, reveals that the pp approximation
is deficient.
It is apparent that the (va)(aa) terms (type IV in Fig.~\ref{fig:diag-phppEKT}) lead to the wrong shape of the dissociation curves. For the N$_{2}$ dimer this contribution is even positive, unlike its negatively-valued ph and PT2 counterparts, 
see the inset in Fig.(\ref{fig:n2}).
Calculations carried out for N$_2$ dimer in an extended active space (see Fig.~S4 in the Supplemental Material) 
indicate that pp approach is flawed also for the (aa)(ao) terms 
(type V in Fig.~\ref{fig:diag-phppEKT})  which deviate strongly from the corresponding ph and PT2 values.

Based on the revealed duality of the multireference ph and pp approximations in the 2nd-order, and using the theoretical and numerical arguments provided above, the  ph-pp combined correlation energy expression reads
\begin{align}
E_{\text{corr}}^{\text{ffAC0}} &  =\frac{1}{2}\sum_{I,J}^{\text{EKT}}\ \left\langle
I|J\right\rangle \ Q_{\mathcal{P}_{\text{ff}}(IJ)}^{\text{ff}}+\frac{1}{2}\sum
_{I,J}^{\text{I,II,IV,V}}\left\langle I|J\right\rangle \ Q_{\mathcal{P}%
_{\text{ph}}(IJ)}^{\text{ph}}\nonumber\\
&  +\frac{1}{2}\sum_{I,J}^{\text{IIIa}}\left\langle I|J\right\rangle
\ Q_{\mathcal{P}_{\text{pp}}(IJ)}^{\text{pp}}%
\ \ \ ,
\label{AC0ff}
\end{align}
where the summations are restricted to terms listed as upper limits.
The combined expression will be referred to as ffAC0. 
Its first term is given as a sum of all 'EKT' terms, identical in  ph and pp approximations. The second and third terms
collect contributions obtained in the ph and pp variants, respectively. 
Note that for a single determinantal reference, only the VII EKT term is present and all three variants: phAC0, ppAC0, and ffAC0 reduce to the  MP2 correlation energy \cite{scuseria2013particle,tahir2019comparing}.

\textit{Results.--}
To compare the performance of the ph and pp approaches and to validate a novel ffAC0 approximation, they are applied to ground and excited states of diverse strength of electron correlation.

Multireference CAS wavefunctions effectively capture static electron correlation 
but their accuracy is limited due to the absence of dynamic correlation.
Dissociation of the fluorine dimer, F$_{2}$, is a striking example of this deficiency, cf.\ Fig.~\ref{fig:f2}. 
The error in dissociation energy  exceeds 20 mHa (see also Table S1 in the Supplemental Material).  
\begin{figure}[h]
    \centering
    \includegraphics[width=0.98\columnwidth]{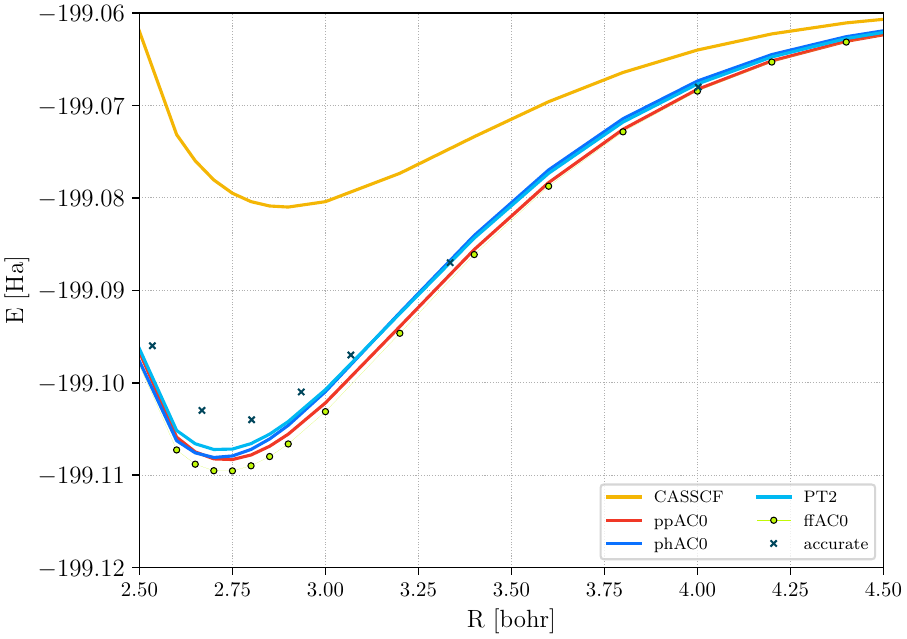}
    \caption{Total energy for the $\ce{F2}$ molecule in cc-pVDZ basis \cite{Dunning:89},  
    calculated with CASSCF(2,2) as a reference wavefunction. 
    All of the curves shifted to match the accurate result for
     R=8 [bohr]. Accurate energies taken from Ref.~\cite{musial2011multi}.} 
    \label{fig:f2}
\end{figure}
All ff correlation methods perform equally well for F$_{2}$ and reduce the error in the dissociation energy of CAS to only 4-5 mHa. 
Systems with more than two correlated electrons pose a challenge for the pp approximation.
On the example of dissociation of H$_{10}$, N$_{2}$, and symmetrically stretched
H$_{2}$O molecules, involving, respectively, 10, 6, and 4 strongly correlated electrons, one observes that the major role for dynamic correlation is
played by terms IV and II. 
Notably, these terms obtained in the 
ph approximation align with their PT2 counterparts, while their values in the pp variant are erroneous (cf.\  Figs.~S1-S4 in the Supplmental Material). 
Consequently, dissociation energy errors (compiled in Table~S1 in the Supplemental Material) 
from the phAC0, ffAC0, and PT2 methods are comparably low, while 
ppAC0 suffers from much higher error, even exceeding 40 mHa for the N$_{2}$ dimer.

\begin{figure}[h]
    \begin{center}
    \includegraphics[width=0.98\columnwidth]{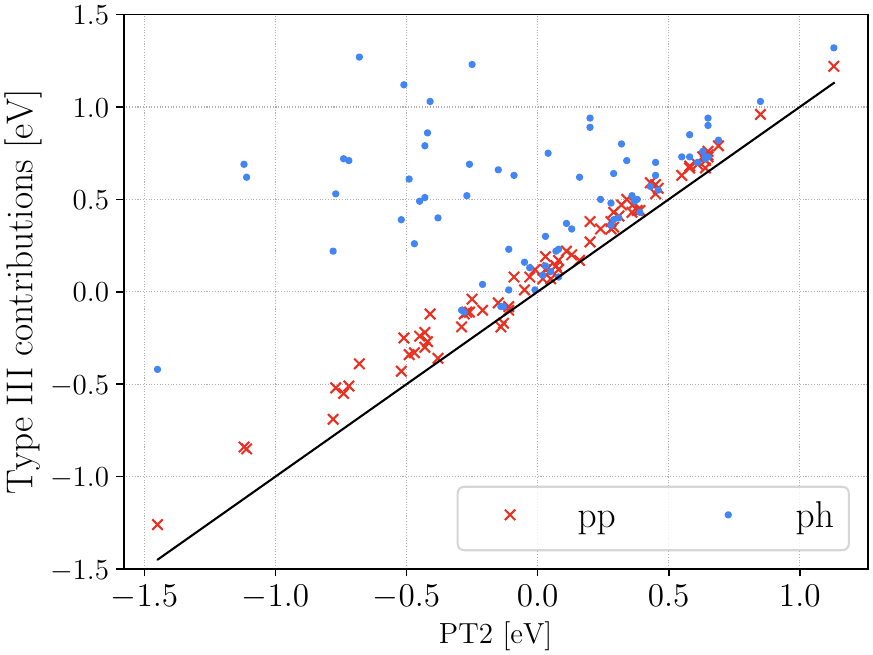}
    \caption{Contributions to excitation energies from the dual terms of type III=IIIa+IIIb (y axis): (vo)(aa)+(va)(ao) in ph and (va)(oa)+ (va)(ao) in pp  versus their PT2 counterparts calculated in TZVP basis \cite{tzvp}.
    Note that term IIIb, identical in both ph and pp approximation, is included here to facilitate direct comparison with PT2, which treats term III as an indivisible entity.}
    \label{fig:correl}
    \end{center}
\end{figure}

Application of the two-fermionic methods to predicting excitation energies shows spectacular performance of the new ffAC0 approach. Recall that ffAC0 differs
from phAC0 by replacing the (vo)(aa)-ph term  by its dual pp counterpart, (va)(oa) 
(type IIIa in Fig.\ref{fig:diag-phppEKT}). The former terms have
been recently identified in the phAC0 method as largely responsible for undercorrelating of excited states \cite{drwal2021,guo2024spinless}. In Fig.~\ref{fig:correl}
we show
that these terms, obtained on a set of valence singlet excitation energies of organic
molecules from the work of Schreiber et al.\ \cite{Thiel:CASPT2} 
indeed deviate strongly from the PT2 equivalent
contributions.
 On the contrary, if those terms are predicted in the pp approximation, the agreement with PT2 is excellent.
 This is due to the fact that excitation
spectrum in the (aa) block of the phEOM lacks negative transitions if $\Psi^{(0)}_0$ describes an excited state. This problem does not exist in the ppEOM which describes electron number changing transitions. 
Thus, the ffAC0 method, where the term IIIa originates from the pp approximation,
is partially free from a deficiency of 
phAC0 and it describes the correlation energy in excited states more accurately than both phAC0 and ppAC0, see Fig.~(\ref{fig:singlets}) and Table~S3 in the Supplemental Material.
As shown in Table~\ref{tab:errors},  
the mean unsigned error (MUE) of ffAC0 amounts to 0.22 eV and it aligns with the error of the PT2 method. 
MUEs of phAC0 and ppAC0 methods are substantially higher, and are equal to  0.55 eV and 0.35 eV, respectively.\cite{gammcor}

\begin{figure}[h]
    \centering
    \includegraphics[width=\textwidth]{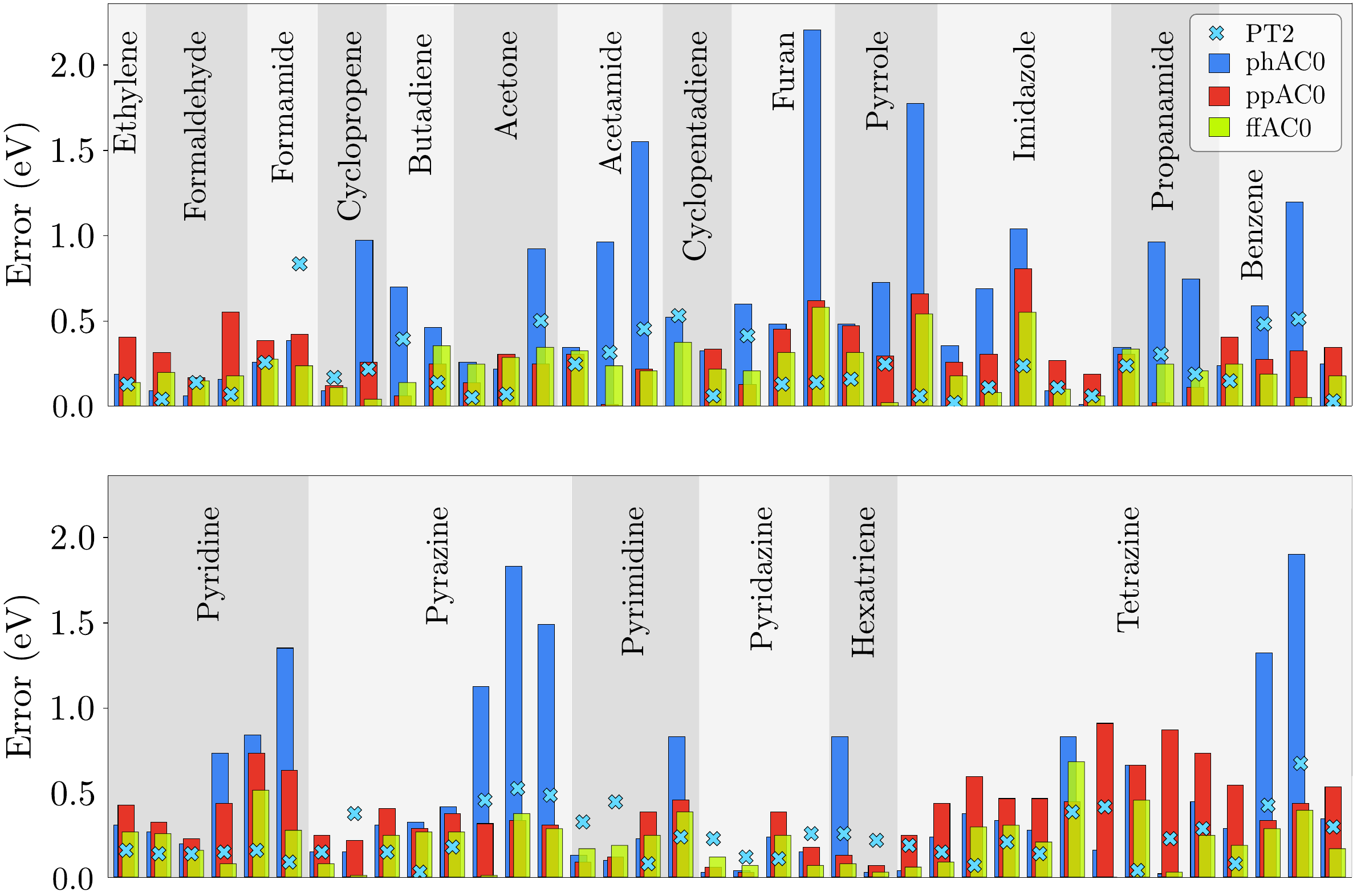}
    \caption{Mean errors in eV for singlet excitations in a set of organic molecules vs.\ CC3 values from Ref.\cite{Thiel:CASPT2}. Calculations done in TZVP basis \cite{tzvp} basis, more computational details are given in the Supplemental Material.
  }
    \label{fig:singlets}
\end{figure}
\begin{table}
  \caption{Mean error (ME), mean unsigned error (MUE) and standard deviation ($\sigma$), in eV, in TZVP basis \cite{tzvp} for the singlet excitations evaluated with respect to CC3 values from Ref.\cite{Thiel:CASPT2}. More computational details are given in the Supplemental Material.
  } 
\begin{center}
\renewcommand{\arraystretch}{1.2}

\centering
 \begin{tabularx}{\columnwidth}
 {G{0.08\columnwidth}
 F{0.16\columnwidth}
 F{0.16\columnwidth}
 F{0.16\columnwidth}
 F{0.16\columnwidth}
 F{0.16\columnwidth}}

  \hline
  &CASSCF&ppAC0&phAC0&ffAC0&PT2\\
  \hline
  ME&0.69&0.32&0.53&0.16 &-0.14 \\
  MUE&0.78&0.35&0.55  & 0.22 & 0.23\\
  $\sigma$&0.59&0.20&0.51 &0.13 &0.17 \\
\end{tabularx}
\end{center}
\label{tab:errors}
\end{table}
Finally, we apply the ff methods to predicting singlet-triplet (ST) splittings in organic biradicals. Their importance has recently increased with rapid advancements in spintronics or photovoltaics technologies \cite{sugawara2011interplay,li2017triplet}. 
Accurate prediction of ST gaps in biradicals is challenging due to the multiconfigurational nature of the singlet state and the need to balance dynamic and static electron correlations in both states. The ppAC0 approach performs exceptionally well, predicting  ST gaps within a 4\% error margin, as shown in Table~\ref{tab:birad}.

\begin{table}
  \caption{ST gaps $(E_T-E_S)$ in [eV] and errors with respect to accurate ('acc.') DEA-EOMCC[4p-2h] results from Ref.~\cite{stoneburner2018mc}. The same geometries as in Ref.~\cite{stoneburner2018mc} are employed.
  Active spaces: 
  CAS(4,4) for $\ce{C4H4}$, (1), $\ce{C4H3NH2}$, (3), $\ce{C4H3CHO}$, (4);
  CAS(4,5) for $\ce{C5H5+}$, (2) and $\ce{C4H2NH2(CHO)}$, (5),
  CAS(6,6) for $\ce{C4H2-1,2-(CH2)2}$, (6) and $\ce{C4H2-1,3-(CH2)2}$, (7), are taken from Ref.~\cite{drwal2022efficient}. 
  All results in the cc-pVDZ basis set, more computational details are given in the Supplemental Material.
  }
\begin{center}
\renewcommand{\arraystretch}{1.2}
\centering
 \begin{tabularx}{\textwidth}
 {G{0.07\textwidth}
 F{0.12\textwidth}
 F{0.12\textwidth}
 F{0.13\textwidth}
 F{0.13\textwidth}
 F{0.14\textwidth}
 F{0.13\textwidth}}
  \hline
  system&CASSCF&ppAC0&phAC0&ffAC0&PT2&acc.\\
  \hline
$(1)$	&0.45	&	0.26	&	0.04	&	0.18	&	-0.08	&0.22\\
$(2)$&	-0.67	&	-0.71	&	-0.88	&	-0.72	&	-0.82	&-0.60\\
$(3)$&	0.39	&	0.15	&	-0.05	&	0.09	&	-0.01	&0.14\\
$(4)$ &	0.42	&	0.19	&	0.02	&	0.13	&	0.05	&0.19\\
$(5)$&	0.29	&	0.27	&	0.19	&	0.28	&	0.17&0.24	\\
$(6)$&	3.33	&	3.59	&	3.48	&	3.57	&	3.39	&3.41\\
$(7)$&	-0.96	&	-0.87	&	-0.98	&	-0.9	&	-0.89	&-0.86\\
\hline
ME	&		0.07	&		0.02		&	-0.13	&		-0.02 	&	 -0.13	&	\\
MUE	&		0.15	&		0.05	&		0.15	&		0.07	&				0.13	&	\\
MU\%E	&		57.69	&		3.56	&		38.16	&		12.41	&			44.14&\\
\end{tabularx}
\end{center}
\label{tab:birad}
\end{table}

Similarly, ffAC0 is capable of predicting correctly both the values, with the error of only 12\%, and signs of the gaps. 
As already reported in Ref.~\cite{drwal2022efficient}, the phAC0 leads to a substantial 40\% error and it struggles to predict correct ordering of the S and T states. PT2  essentially follows the behavior of phAC0, leading to similar errors.

\textit{Concluding remarks.--}
We have presented a unified formulation of a multireference two-fermionic approach to the correlation energy, which is consistent with the many-electron perturbation theory in the 2nd order. The revealed duality between ph and pp contributions to the correlation energy has opened a way for combining the pp and ph terms in a non-trivial way.
Notice that so far theoretical approaches involving combinations of ph and pp channels have been based on a single reference model in the 0th order \cite{scuseria2013particle,tahir2019comparing,kaufmann2021self}.
The proposed combined method, ffAC0, inferred from theoretical and numerical arguments, overall surpasses in performance not only pp and ph variants but also the 2nd-order perturbation method NEVPT2. 
It achieves similar accuracy for ground and singlet excited states as PT2 but it is superior to the latter in predicting ST splittings in biradicals. This is a remarkable result, taking into account that ffAC0 relies on only 1- and 2-RDMs.
It can then treat large active spaces, unattainable for available multireference PT methods.

The presented technique that has led to unifying multireference ph and pp correlation energies in the 2nd-order can be extended to higher orders, possibly leading to improved accuracy. 

This work has been supported by the National Science Center of Poland under grant no.\ 2019/35/B/ST4/01310, and the National Natural Science Foundation of China (Grant No. 22273052). 

\bibliography{biblio}

\end{document}


\title{Duality of particle-hole and particle-particle theories for
strongly correlated electronic systems\\
Supplemental Material}
\author{Aleksandra Tucholska}
\affiliation{Institute of Physics, Lodz University of Technology,  ul.\ Wolczanska 217/221, 93-005 Lodz, Poland}
\author{Yang Guo}
\affiliation{Qingdao Institute for Theoretical and Computational Sciences, Institute of Frontier Chemistry, School of Chemistry and Chemical Engineering, Shandong University, Qingdao, Shandong 266237}
\author{Katarzyna Pernal}
\email{pernalk@gmail.com}
\affiliation{Institute of Physics, Lodz University of Technology,  ul.\ Wolczanska 217/221, 93-005 Lodz, Poland}
\date{\today}

\maketitle
\tableofcontents
\section{Extended RPA  equations}\label{SI-sec:equations}
As a result of assuming the killer condition, see Refs.~\cite{rowe, erpa1}, the main matrix of the EOM in the multireference (extended) RPA approximation, Eq.(12) in the main text, is given in terms of 1- and 2-RDMs.
Explicit expression for the ph matrix can be found in Ref. \cite{erpa1}. For the pp extended-RPA, the main matrix is shown for the first time and it reads
\begin{equation}
    A_{IJ}^\alpha = A_{pqrs}^\alpha = \left\langle \Psi_{0}^{(0)}\right\vert \left[  
 \hat{o}_{J},
 [\hat{H}^{\alpha},
 \hat{o}_{I}^{\dagger}]\right]  \left\vert \Psi_{0}^{(0)}%
\right\rangle 
=\left\langle \Psi_{0}^{(0)}\right\vert \left[  
\hat{a}_r\hat{a}_s,
[\hat{H}^{\alpha},
\hat{a}_p^{\dagger}\hat{a}_q^{\dagger}]\right]  \left\vert \Psi_{0}^{(0)}%
\right\rangle \ \ \ .
\end{equation}
The expression for the matrix element $A_{pqrs}^{\alpha}$, with general indices related to spinorbitals, is explicitly defined using one- and two-electron reduced density matrices, $\gamma$ and $\Gamma$, as follows
  \begin{equation}\begin{split}
    &A_{pq rs}^\alpha
=\tilde{g}_{pqrs}^\alpha
  +h_{qs}^{\alpha}\delta_{pr}
-h_{qr}^{\alpha}\delta_{ps}
-h_{ps}^{\alpha}\delta_{qr}
+h_{pr}^{\alpha}\delta_{qs}\\&
-\gamma_{sq}h_{pr}^{\alpha}
+\gamma_{rq}h_{ps}^{\alpha}
+\gamma_{sp}h_{qr}^{\alpha}
-\gamma_{rp}h_{qs}^{\alpha}\\&
-\frac{1}{2}\sum_{t}\gamma_{tq}h_{st}^{\alpha}\delta_{pr}
-\frac{1}{2}\sum_{t}\gamma_{st}h_{qt}^{\alpha}\delta_{pr}
+\frac{1}{2}\sum_{t}\gamma_{tq}h_{rt}^{\alpha}\delta_{ps}
+\frac{1}{2}\sum_{t}\gamma_{rt}h_{qt}^{\alpha}\delta_{ps}\\&
+\frac{1}{2}\sum_{t}\gamma_{st}h_{pt}^{\alpha}\delta_{qr}
+\frac{1}{2}\sum_{t}\gamma_{tp}h_{st}^{\alpha}\delta_{qr}
-\frac{1}{2}\sum_{t}\gamma_{tp}h_{rt}^{\alpha}\delta_{qs}
-\frac{1}{2}\sum_{t}\gamma_{rt}h_{pt}^{\alpha}\delta_{qs}\\&
+\sum_{t}\gamma_{st}\tilde{g}^\alpha_{ pqtr}
+\sum_{t}\gamma_{rt}\tilde{g}^\alpha_{ pqst}
+\sum_{t}\gamma_{tp}\tilde{g}^\alpha_{ qtrs}
+\sum_{t}\gamma_{tq}\tilde{g}^\alpha_{ ptsr}\\&
+\sum_{tu}\gamma_{tu}\tilde{g}^\alpha_{ qtsu}\delta_{pr}
+\sum_{tu}\gamma_{tu}\tilde{g}^\alpha_{ qtur}\delta_{ps}
+\sum_{tu}\gamma_{tu}\tilde{g}^\alpha_{ ptus}\delta_{qr}
+\sum_{tu}\gamma_{tu}\tilde{g}^\alpha_{ ptru}\delta_{qs}\\&
+\sum_{tu}\Gamma_{stqu}\tilde{g}^\alpha_{ ptur}
+\sum_{tu}\Gamma_{stpu}\tilde{g}^\alpha_{ qtru}
+\sum_{tu}\Gamma_{rtpu}\tilde{g}^\alpha_{ qtus}
+\sum_{tu}\Gamma_{rtqu}\tilde{g}^\alpha_{ ptsu}\\&
+\frac{1}{4}\sum_{tuv}\Gamma_{tuqv}\big(\tilde{g}^\alpha_{ svut}\delta_{pr} + \tilde{g}^\alpha_{ rvtu}\delta_{ps}\big)
+\frac{1}{4}\sum_{tuv}\Gamma_{stuv}\big(\tilde{g}^\alpha_{ qtvu}\delta_{pr}+ \tilde{g}^\alpha_{ ptuv}\delta_{qr}\big)\\&
+\frac{1}{4}\sum_{tuv}\Gamma_{rtuv}\big(\tilde{g}^\alpha_{ qtuv}\delta_{ps}+ \tilde{g}^\alpha_{ ptvu}\delta_{qs}\big)
+\frac{1}{4}\sum_{tuv}\Gamma_{tupv}\big(\tilde{g}^\alpha_{ svtu}\delta_{qr}+ \tilde{g}^\alpha_{ rvut}\delta_{qs}\big) \ \ \ .
 \end{split}\end{equation}
The 1- and 2-RDMs correspond to a given reference wavefunction $\Psi^{(0)}_0$ {and they are defined as
 \begin{equation}
 \begin{split}
    &\gamma_{pq} =\left\langle \Psi_{0}^{(0)}|\hat{a}_{q}^{\dagger}\hat{a}_{p}|\Psi_{0}^{(0)}\right\rangle
    \\&\Gamma_{pqrs} =\left\langle \Psi_{0}^{(0)}|\hat{a}_{r}^{\dagger}\hat{a}_{s}^{\dagger}\hat{a}_{q}\hat{a}_{p}|\Psi_{0}^{(0)}\right\rangle     \ \ \ .
 \end{split}
 \end{equation}
$\tilde{g}_{pqrs}^\alpha$ is a symmetrized product of an $\alpha$-dependent factor and a two-electron integral $\langle pq| rs \rangle$, given as follows
 \begin{equation}
     \tilde{g}_{pqrs}^\alpha = g_{pqrs}^\alpha - g_{pqsr}^\alpha
 \end{equation}
with
\begin{equation}
g^{\alpha}_{pqrs} = \delta_{I_{p} I_{q}} \delta_{I_{q} I_{r}} \delta_{I_{r} I_{s}} \delta_{I_{p},1} \langle pq | rs \rangle
+\alpha(1-\delta_{I_{p} I_{q}} \delta_{I_{q} I_{r}} \delta_{I_{r} I_{s}} \delta_{I_{p},1}) \langle pq | rs \rangle \ \ \ ,
\end{equation}
Modified one-electron Hamiltonian elements read
\begin{eqnarray}
&
h^{\alpha}_{pq} = \alpha h_{pq} + (1-\alpha)\delta_{I_{p} I_{q}} h_{pq}^{\text{eff}}\\&
h_{pq}^{\text{eff}}=h_{pq}+\sum_{rs}(1-\delta_{I_{p}I_{r}}\delta_{I_{p},1})\gamma_{rs}\left[  \left\langle
pr|qs\right\rangle -\left\langle pr|sq\right\rangle \right] \ \ \ .
\end{eqnarray}
$I_p$ denotes a set of doubly occupied, active or virtual spinorbitals that an orbital $p$ belongs to. $I=1$ is used for a set of active orbitals,  thus $\delta_{I_p,1}=1$
denotes that $p$ is an active orbital.
Two electron integrals are given in physicist notation, namely
\begin{equation}
    \langle pq|rs  \rangle = \langle \chi_p(1)\chi_q(2) |  r_{12}^{-1}   | \chi_r(1)\chi_s(2) \rangle \ \ \ .
\end{equation}

\section{Unique permutations in $Q^{\rm{ff}}$ matrices}\label{SI-sec:corresp}
The summation over $i$ in $Q_{pqrs}^{\text{ff}}$ expression (see Eq.(16) in the main text)
\begin{equation}
Q_{pqrs}^{\text{ff}}= \sum_i \sum_{\mu,\nu \in {\cal H}_M}  \left[  \mathbf{\gamma}_{\text{ff}}^{\mu
}\right]  _{p_iq_i}^{(0)} D_{\mu\nu}^{\text{ff}}\left[  \mathbf{\gamma}_{\text{ff}%
}^{\nu}\right]  _{r_is_i}^{(0)}\ \ \ ,\label{QIJ}%
\end{equation}
denotes all unique permutations $p\leftrightarrow q$, $r\leftrightarrow s$, and $pq\leftrightarrow rs$ which do not change a two-electron integral that multiplies $Q^{\rm ff}_{pqrs}$, (recall that  $Q^{\rm ph}_{pqrs}$ is multiplied by $\langle pr| qs\rangle$, while $Q^{\rm pp}_{pqrs}$ by $\langle pq |rs\rangle$). On the example of terms IIIa and IIIb we show all of the unique permutations taken into account. 
The contributions to the IIIa term arise from multiplying the integral type $\langle\text{va}|\text{oa}\rangle$ with distinct permutations (vo)(aa), (aa)(vo), (aa)(ov), and (ov)(aa) for the ph variant, and (va)(oa), (av)(ao), (ao)(av), and (oa)(va) for the pp variant. For the IIIb term, contributions stem from the integral type $\langle\text{va}|\text{ao}\rangle$, associated with unique permutations (va)(ao), (ao)(va), (av)(oa), (oa)(av), (av)(ao), (ao)(av), (va)(oa), and (oa)(va) for the ph variant, and (va)(ao), (av)(oa), (ao)(va), and (oa)(av) for the pp variant. 




\section{Ground states}\label{SI-sec:ground}
All CASSCF calculations were performed in the Dalton program \cite{aidas2014d, Dalton2022}. 
All AC methods were implemented in the GammCor program \cite{gammcor}.  We employed cc-pVDZ basis for all of the potential energy surface calculations. 
PT2 indicates PC-NEVPT2 computed using Dalton program \cite{aidas2014d, Dalton2022}.




\addcontentsline{toc}{subsection}{\protect\numberline{\ref{SI-fig:h2o-44-contr}}
{$\ce{H2O}$ CAS(4,4), contributions I-V plot. }}
\begin{figure}[h]
    \centering
    \includegraphics[width=0.6\textwidth]{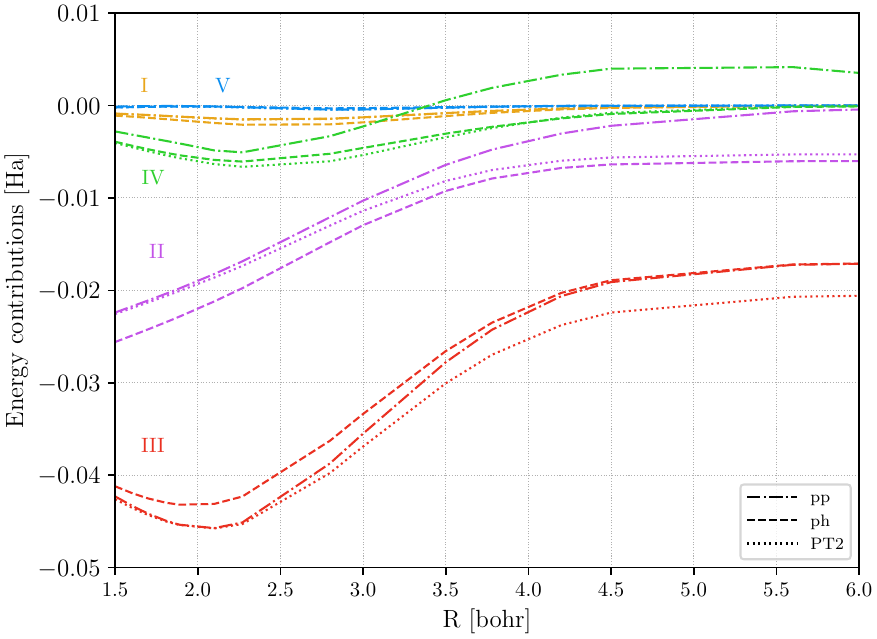}
    \caption{Energy contributions for $\ce{H2O}$ calculated with CASSCF(4,4) as a reference. Dashdotted lines  - pp contributions, dashed lines - ph contributions, dotted lines PT2 contributions. EKT terms are not shown.}
    \label{SI-fig:h2o-44-contr}
\end{figure}

\addcontentsline{toc}{subsection}{\protect\numberline{\ref{SI-fig:h2o-88-contr}}
{$\ce{H2O}$ CAS(8,8) Contributions I-V plot.}}
\begin{figure}[h]
    \centering
    \includegraphics[width=0.6\textwidth]{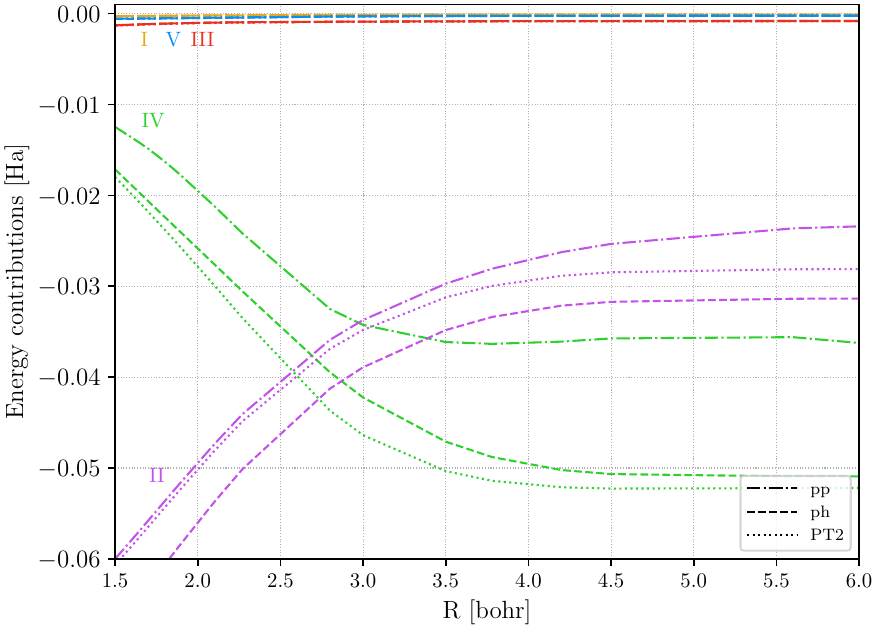}
    \caption{Energy contributions for $\ce{H2O}$ calculated with CASSCF(8,8) as a reference. Dashdotted lines  - pp contributions, dashed lines - ph contributions, dotted lines PT2 contributions. EKT terms are not shown.}
    \label{SI-fig:h2o-88-contr}
\end{figure}

\addcontentsline{toc}{subsection}{\protect\numberline{\ref{SI-fig:n2-66-contr}}
{$\ce{N2}$ CAS(6,6) Contributions I-V plot.}}

\begin{figure}[h]
    \centering
    \includegraphics[width=0.6\textwidth]{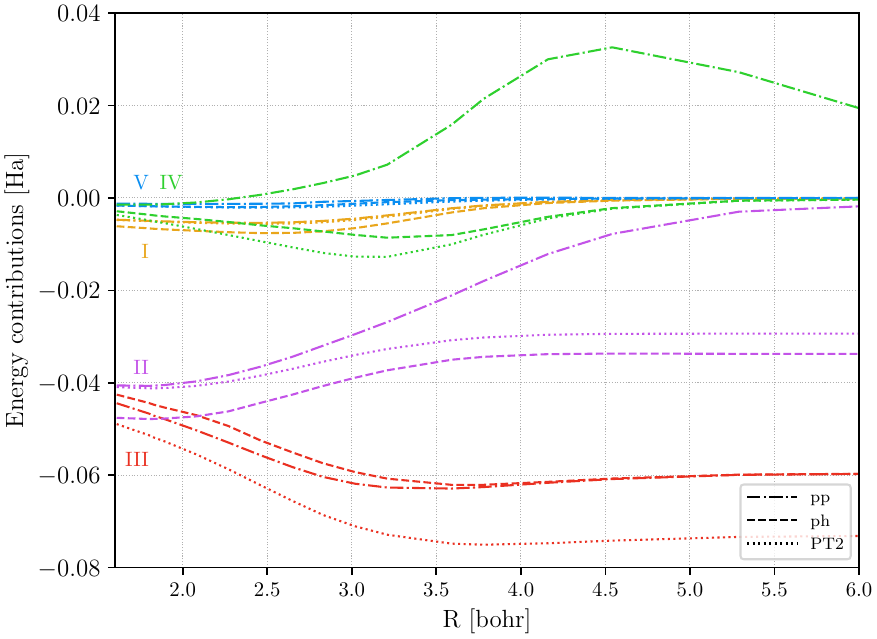}
    \caption{Energy contributions for $\ce{N2}$ calculated with CASSCF(6,6) as a reference. Dashdotted lines  - pp contributions, dashed lines - ph contributions, dotted lines PT2 contributions. EKT terms are not shown.}
    \label{SI-fig:n2-66-contr}
\end{figure}


\addcontentsline{toc}{subsection}{\protect\numberline{\ref{SI-fig:n2-622-contr}}
{$\ce{N2}$ CAS(6,22) Contributions I-V plot.}}
\begin{figure}[h]
    \centering
    \includegraphics[width=0.6\textwidth]{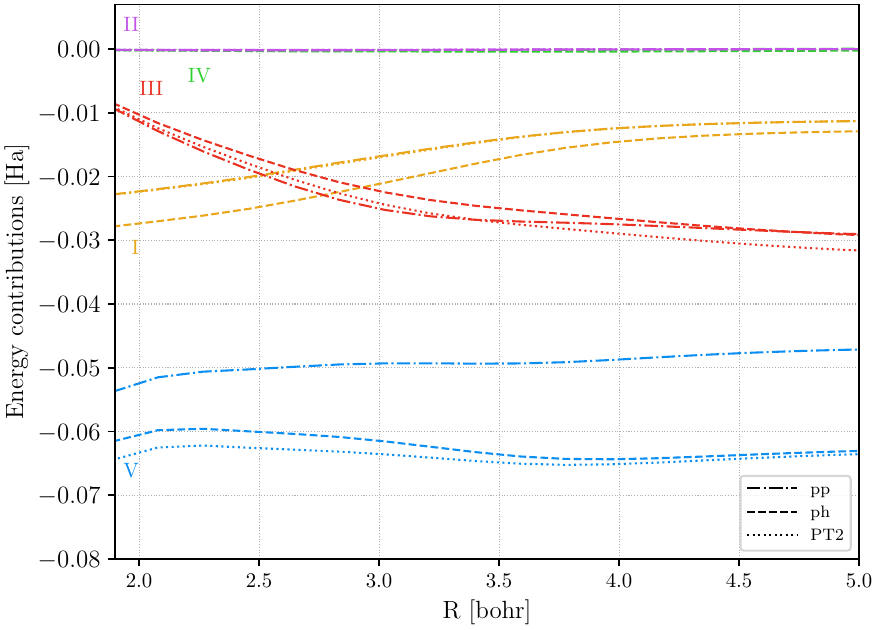}
    \caption{Energy contributions for $\ce{N2}$ calculated with CASSCF(6,22) as a reference. Dashdotted lines  - pp contributions, dashed lines - ph contributions, dotted lines PT2 contributions. EKT terms are not shown.}
    \label{SI-fig:n2-622-contr}
\end{figure}

\addcontentsline{toc}{subsection}{\protect\numberline{\ref{SI-tab:diss}}
{Table of dissociation energies and errors.}}
\begin{table}
  \caption{  Equilibrium geometries are 
  R$_{\text{H-H}}$=1.8 [bohr] for $\ce{H10}$, 
 R=2.8 [bohr] for $\ce{F2}$, 
 R$_{\text{O-H}}$=1.8 [bohr] for $\ce{H2O}$, and
 R=2.08 [bohr] for $\ce{N2}$.
  Accurate  energies for equilibrium geometry 
  are  $\text{acc.} = -5.614 $ [Ha] \cite{eriksen2019many} for $\ce{H10}$, 
  $\text{acc.} =-199.104$ [Ha] \cite{musial2011multi} for $\ce{F2}$,
  $\text{acc.} =-76.241$ [Ha] \cite{eriksen2019many} for $\ce{H2O}$, and 
  $\text{acc.} = -109.2773$[Ha] \cite{eriksen2019many} for $\ce{N2}$.
  Equilibrium geometries in [Ha], dissociation energies and errors in [mHa]. PT2 indicates PC-NEVPT2 computed using Dalton program \cite{aidas2014d, Dalton2022}.}
\begin{center}
\renewcommand{\arraystretch}{0.7}
\centering
 \begin{tabularx}{\textwidth}
 {G{0.1\textwidth}
 F{0.08\textwidth}
 F{0.08\textwidth}
 F{0.08\textwidth}
 F{0.08\textwidth}
 F{0.08\textwidth}
 F{0.08\textwidth}
 F{0.08\textwidth}
 F{0.08\textwidth}
 F{0.08\textwidth}
 F{0.08\textwidth}}
  \multicolumn{6}{c}{\scriptsize{Energy at equilibrium geometry }}
 &\multicolumn{5}{c}{\scriptsize{Dissociation energies (error vs acc.)}}\\
   \noalign{\global\arrayrulewidth=0.7pt}
   \hline
ref.	&	CASSCF	&	ppAC0	&	phAC0	&	ffAC0	&	PT2	&	CASSCF	&	ppAC0	&	phAC0	&	ffAC0	&	PT2\\
\noalign{\global\arrayrulewidth=0.07pt}
\hline
 \multicolumn{11}{c}{\scriptsize{$\ce{H10}$ }}\\
   \noalign{\global\arrayrulewidth=0.05pt}
\hline
CAS(10,10)	&	-5.5115	&	-5.5784	&	-5.5920	&	-5.5920	&	-5.5879	&	519.6	&	585.6	&	599.2	&	599.2	&	595.1\\
 &&&&& &	(-90.7)	&	(-24.7)	&	(-11.1)	&	(-11.1)	&	(-15.2)\\
\noalign{\global\arrayrulewidth=0.7pt}
   \hline
    \multicolumn{11}{c}{\scriptsize{$\ce{F2}$  }}\\
     \noalign{\global\arrayrulewidth=0.05pt}  
      \hline
      CAS(2,2)	&	-198.7651	&	-199.0827	&	-199.0821	&	-199.0838	&	-199.0840	&	21.9	&	48.8	&	48.9	&	50.2	&	47.9\\
&&&&&&	(-23.1)	&	(3.8)	&	(3.9)	&	(5.2)	&	(2.9 )  \\
           \noalign{\global\arrayrulewidth=0.05pt}  \hline
    \multicolumn{11}{c}{\scriptsize{$\ce{H2O}$}}\\
      \hline
      CAS(4,4)	&	-76.0776	&	-76.2254	&	-76.2283	&	-76.2302	&	-76.2274	&	291.7	&	341.1	&	336.1	&	338.0	&	332.4\\
&&&&&&	(-41.0)	&	(8.5)	&	(3.4)	&	(5.4)	&	(-0.2)\\
CAS(8,8)	&	-76.1472	&	-76.2194	&	-76.2325	&	-76.2325	&	-76.2276	&	332.0	&	341.4	&	333.2	&	333.2	&	330.3\\
&&&&&&	(-0.7)	&	(8.8)	&	(0.6)	&	(0.6)	&	(-2.3)\\
            \noalign{\global\arrayrulewidth=0.05pt}\hline
    \multicolumn{11}{c}{\scriptsize{$\ce{N2}$} }\\
   \noalign{\global\arrayrulewidth=0.05pt}   \hline
CAS(6,6)	&	-109.0902	&	-109.2349	&	-109.2442	&	-109.2466	&	-109.2476	&	313.7	&	358.6	&	325.6	&	328.1	&	320.7\\
&&&&&&	(-4.5)	&	(40.4)	&	(7.4)	&	(9.9)	&	(2.4)\\
\end{tabularx}
\end{center}
\label{SI-tab:diss}
\end{table}

\addcontentsline{toc}{subsection}{\protect\numberline{\ref{SI-tab:contrib}}
{Table of numerical values for contributions I-V.}}
\begin{table}
  \caption{Energy contributions  calculated with
  ppAC0, phAC0 and PT2 in cc-pVDZ basis set. Contributions given in [Ha].
  For $\ce{N2}$ we used CASSCF(6,6) reference wavefunction, for $\ce{H2O}$ we used CASSCF(4,4) reference wavefunction,
  and for $\ce{F2}$ we used CAS(2,2) wavefunction. EKT terms are IIIb, VI, VII, VIII.
  PT2 indicates PC-NEVPT2 computed using Dalton program \cite{aidas2014d, Dalton2022}.  }
\begin{center}
\renewcommand{\arraystretch}{0.7}
\centering
 \begin{tabularx}{\textwidth}
 {G{0.03\textwidth}
 F{0.08\textwidth}
 F{0.085\textwidth}
 F{0.085\textwidth}
 F{0.085\textwidth}
 F{0.085\textwidth}
 F{0.085\textwidth}
 F{0.085\textwidth}
 F{0.085\textwidth}
 F{0.085\textwidth}
 F{0.085\textwidth}}
 \noalign{\global\arrayrulewidth=0.7pt}\hline
 \multicolumn{11}{c}{\scriptsize{ppAC0}}\\
   \noalign{\global\arrayrulewidth=0.05pt}
   \hline
 	&	R [bohr]	&	I	&	II	&	IIIa	&	IIIb	&	IV	&	V 	&	VI	&	VII	&	VIII	\\
\hline
$\ce{N2}$	&	2.08	&	-0.0053	&	-0.0397	&	-0.0176	&	-0.0327	&	-0.0010	&	-0.0013	&	-0.0067	&	-0.0174	&	-0.0231	\\
	&	5.29	&	-0.0001	&	-0.0030	&	-0.0004	&	-0.0595	&	0.0272	&	0.0000	&	-0.0035	&	-0.0140	&	-0.0310	\\
$\ce{H2O}$	&	1.81	&	-0.0011	&	-0.0204	&	-0.0399	&	-0.0051	&	-0.0038	&	0.0000	&	-0.0055	&	-0.0399	&	-0.0321	\\
	&	6.50	&	0.0000	&	-0.0002	&	-0.0001	&	-0.0169	&	0.0027	&	0.0000	&	-0.0100	&	-0.0344	&	-0.0388	\\
$\ce{F2}$	&	2.80	&	-0.0023	&	-0.0023	&	-0.0372	&	-0.0100	&	0.0000	&	0.0000	&	-0.0216	&	-0.1847	&	-0.0595	\\
	&	8.00	&	0.0000	&	0.0000	&	0.0000	&	-0.0134	&	0.0000	&	0.0000	&	-0.0193	&	-0.1878	&	-0.0697	\\
\noalign{\global\arrayrulewidth=0.7pt} \hline
 \multicolumn{11}{c}{\scriptsize{phAC0}}\\
   \noalign{\global\arrayrulewidth=0.05pt}
   \hline
    	&	R [bohr]	&	I	&	II	&	IIIa	&	IIIb	&	IV	&	V	&	VI	&	VII	&	VIII	\\
    \hline
   $\ce{N2}$	&	2.08	&	-0.0072	&	-0.0471	&	-0.0148	&	-0.0327	&	-0.0047	&	-0.0019	&	-0.0067	&	-0.0174	&	-0.0231	\\
	&	5.29	&	-0.0002	&	-0.0337	&	-0.0004	&	-0.0595	&	-0.0006	&	0.0000	&	-0.0035	&	-0.0140	&	-0.0310	\\
$\ce{H2O}$	&	1.81	&	-0.0015	&	-0.0234	&	-0.0379	&	-0.0051	&	-0.0051	&	-0.0001	&	-0.0055	&	-0.0399	&	-0.0321	\\
	&	6.50	&	0.0000	&	-0.0060	&	-0.0001	&	-0.0169	&	0.0000	&	0.0000	&	-0.0100	&	-0.0344	&	-0.0388	\\
$\ce{F2}$	&	2.80	&	-0.0032	&	-0.0026	&	-0.0354	&	-0.0100	&	0.0000	&	0.0000	&	-0.0216	&	-0.1847	&	-0.0595	\\
	&	8.00	&	0.0000	&	0.0000	&	0.0000	&	-0.0134	&	0.0000	&	0.0000	&	-0.0193	&	-0.1878	&	-0.0697	\\
\noalign{\global\arrayrulewidth=0.7pt}  \hline
 \multicolumn{11}{c}{\scriptsize{PT2}}\\
   \noalign{\global\arrayrulewidth=0.05pt}
   \hline
	&	R [bohr]	&	I	&	II	&	\multicolumn{2}{c}{\scriptsize{IIIa+IIIb}}			&	IV	&	V	&	VI	&	VII	&	VIII	\\
\hline
   $\ce{N2}$	&	2.08	&	-0.0054	&	-0.0407	&	\multicolumn{2}{c}{\scriptsize{-0.0555}} &	-0.0067	&	-0.0020	&	-0.0067	&	-0.0174	&	-0.0231	\\
	&	5.29	&	-0.0002	&	-0.0294	&	\multicolumn{2}{c}{\scriptsize{-0.0734}}	    &	-0.0006	&	0.0000	&	-0.0035	&	-0.0140	&	-0.0310	\\
$\ce{H2O}$	&	1.81	&	-0.0011	&	-0.0206	&	\multicolumn{2}{c}{\scriptsize{-0.0450}}	&	-0.0054	&	-0.0001	&	-0.0055	&	-0.0399	&	-0.0321	\\
	&	6.50	&	0.0000	&	-0.0053	&	\multicolumn{2}{c}{\scriptsize{-0.0205}}	    &	0.0000	&	0.0000	&	-0.0100	&	-0.0344	&	-0.0388	\\
$\ce{F2}$	&	2.80	&	-0.0023	&	-0.0023	&	\multicolumn{2}{c}{\scriptsize{-0.0485}}	&	0.0000	&	0.0000	&	-0.0216	&	-0.1847	&	-0.0595	\\
	&	8.00	&	0.0000	&	0.0000	&	\multicolumn{2}{c}{\scriptsize{-0.0159}}	    &	0.0000	&	0.0000	&	-0.0193	&	-0.1878	&	-0.0697	\\
\end{tabularx}
\end{center}
\label{SI-tab:contrib}
\end{table}

\section{Excited states}\label{SI-sec:excit}
All CASSCF calculations were performed in the Dalton program \cite{aidas2014d, Dalton2022}, and are SS-CAS calculations.
All AC methods were implemented in the GammCor program \cite{gammcor}.  We employed the cc-pVDZ basis for ST gaps in biradicals and the TZVP basis \cite{tzvp} for vertical excitation energies.
PT2 indicates PC-NEVPT2 computed using Dalton program \cite{aidas2014d, Dalton2022}.

\clearpage
\setlength{\LTcapwidth}{\textwidth}
\addcontentsline{toc}{subsection}{\protect\numberline{\ref{SI-tab:all-exc}}
{Table of vertical excitation energies.}}
\begin{center}
\renewcommand{\arraystretch}{0.7}
\begin{longtable}
{G{0.15\textwidth}  
G{0.15\textwidth}
F{0.1\textwidth}
F{0.1\textwidth}
F{0.1\textwidth}
F{0.1\textwidth}
F{0.1\textwidth}
F{0.1\textwidth}} 
\caption{Vertical excitation energies for singlet excitations. 
Geometries of molecules (MP2/6-31G*) have been taken from \cite{Thiel:CASPT2}.
The active spaces of the SS-CAS calculations are the same as those in the work of Schreiber et al.\ \cite{Thiel:CASPT2}.
All calculations of excited states have been performed in the standard for the
considered excitation TZVP basis set \cite{tzvp}. All values are in [eV].
}\\
\hline
Molecule	&	State		&	CASSCF	&	ppAC0	&	phAC0	&	ffAC0	&	PT2	&	CC3 \cite{Thiel:08}	\\
 \hline \hline
\endfirsthead

\multicolumn{8}{c}%
{{ \tablename\ \thetable{} -- continued from previous page}} \\
\hline
Molecule	&	State		&	CASSCF	&	ppAC0	&	phAC0	&	ffAC0	&	PT2	&	CC3	\\
 \hline \hline
\endhead

\hline \multicolumn{8}{r}{{Continued on next page}} \\ 
\endfoot

\hline \hline
\endlastfoot
 Ethylene	&	$\sBju$	($\pi\rightarrow \pi^*$)	&	9.13	&	8.78	&	8.56	&	8.23	&	8.24	&	8.37	\\
  \hline
  Formaldehyde	&	1 $\sAd$	($n\rightarrow \pi^*$)	&	5.14	&	4.27	&	4.04	&	4.15	&	3.99	&	3.95	\\
	&	1 $\sBj$	($\sigma\rightarrow \pi^*$)	&	10.49	&	9.35	&	9.24	&	9.33	&	9.04	&	9.18	\\
	&	2 $\sAj$	($\pi\rightarrow \pi^*$)	&	10.76	&	10	&	9.28	&	9.26	&	9.51	&	9.44	\\
   \hline
   Formamide	&	1 $\sAb$	($n\rightarrow \pi^*$)	&	5.41	&	6.04	&	5.91	&	5.93	&	5.91	&	5.65	\\
	&	2 $\sAp$	($\pi\rightarrow \pi^*$)	&	8.93	&	7.84	&	8.66	&	8.03	&	7.42	&	8.27	\\
   \hline
   Cyclopropene	&	1 $\sBj$	($\sigma\rightarrow \pi^*$)	&	7.27	&	7.02	&	6.81	&	6.79	&	6.73	&	6.9	\\
	&	1 $\sBd$	($\pi\rightarrow \pi^*$)	&	8.48	&	7.36	&	8.09	&	7.14	&	6.88	&	7.1	\\
   \hline
 E-Butadiene	&	$\sBju$	($\pi\rightarrow \pi^*$)	&	7.22	&	6.64	&	7.29	&	6.44	&	6.18	&	6.58	\\
	&	2 $\sAg$	($\pi\rightarrow \pi^*$)	&	5.75	&	7.02	&	7.24	&	7.13	&	6.91	&	6.77	\\
   \hline
   Acetone	&	1 $\sAd$	($n\rightarrow \pi^*$)	&	4.47	&	4.54	&	4.66	&	4.65	&	4.45	&	4.4	\\
	&	1 $\sBj$	($\sigma\rightarrow \pi^*$)	&	9.32	&	9.48	&	9.39	&	9.46	&	9.1	&	9.17	\\
	&	2 $\sAj$	($\pi\rightarrow \pi^*$)	&	10.81	&	9.4	&	10.59	&	10	&	9.14	&	9.65	\\
   \hline
Acetamide	&	1 $\sAb$	($n\rightarrow \pi^*$)	&	5.51	&	6	&	6.04	&	6.02	&	5.94	&	5.69	\\
	&	2 $\sAp$	($\pi\rightarrow \pi^*$)	&	8.82	&	7.68	&	8.65	&	7.91	&	7.35	&	7.67	\\
	&	3 $\sAp$	($\pi\rightarrow \pi^*$)	&	12.09	&	10.72	&	12.08	&	10.71	&	10.04	&	10.5	\\
   \hline
Cyclopentadiene	&	1 $\sBd$	($\pi\rightarrow \pi^*$)	&	7.28	&	5.73	&	6.26	&	5.35	&	5.19	&	5.73	\\
	&	2 $\sAj$	($\pi\rightarrow \pi^*$)	&	6.61	&	6.95	&	6.94	&	6.83	&	6.67	&	6.61	\\
   \hline
Furan	&	1 $\sBd$	($\pi\rightarrow \pi^*$)	&	7.88	&	6.73	&	7.21	&	6.39	&	6.18	&	6.6	\\
	&	2 $\sAj$	($\pi\rightarrow \pi^*$)	&	6.74	&	7.08	&	7.11	&	6.94	&	6.75	&	6.62	\\
	&	3 $\sAj$	($\pi\rightarrow \pi^*$)	&	10.11	&	9.16	&	10.78	&	9.12	&	8.39	&	8.53	\\
   \hline
Pyrrole	&	2 $\sAj$	($\pi\rightarrow \pi^*$)	&	6.54	&	6.88	&	6.89	&	6.72	&	6.56	&	6.4	\\
	&	1 $\sBd$	($\pi\rightarrow \pi^*$)	&	7.74	&	7.01	&	7.45	&	6.73	&	6.46	&	6.71	\\
	&	3 $\sAj$	($\pi\rightarrow \pi^*$)	&	9.52	&	8.84	&	9.98	&	8.72	&	8.11	&	8.17	\\
   \hline
Imidazole	&	2 $\sAp$	($\pi\rightarrow \pi^*$)	&	6.87	&	6.84	&	6.94	&	6.76	&	6.6	&	6.58	\\
	&	3 $\sAp$	($\pi\rightarrow \pi^*$)	&	7.87	&	7.41	&	7.8	&	7.18	&	6.99	&	7.1	\\
	&	4 $\sAp$	($\pi\rightarrow \pi^*$)	&	9.49	&	9.27	&	9.51	&	9.01	&	8.69	&	8.45	\\
	&	1 $\sAb$	($n\rightarrow \pi^*$)	&	6.76	&	7.09	&	6.91	&	6.92	&	6.93	&	6.82	\\
	&	2 $\sAb$	($\pi\rightarrow \pi^*$)	&	8.15	&	8.12	&	7.92	&	7.87	&	7.87	&	7.93	\\
   \hline
   Propanamide	&	1 $\sAb$	($n\rightarrow \pi^*$)	&	5.54	&	6.03	&	6.07	&	6.06	&	5.96	&	5.72	\\
	&	2 $\sAp$	($\pi\rightarrow \pi^*$)	&	8.79	&	7.64	&	8.6	&	7.87	&	7.31	&	7.62	\\
	&	3 $\sAp$	($\pi\rightarrow \pi^*$)	&	11.77	&	10.17	&	10.82	&	10.27	&	9.87	&	10.06	\\
   \hline
   Benzene	&	1 $\sBdu$ 	($\pi\rightarrow \pi^*$)	&	4.98	&	5.48	&	5.31	&	5.32	&	5.22	&	5.07	\\
	&	1 $\sBju$	($\pi\rightarrow \pi^*$)	&	7.87	&	6.96	&	7.28	&	6.49	&	6.19	&	6.68	\\
	&	1 $\sEju$ 	($\pi\rightarrow \pi^*$)	&	9.21	&	7.78	&	8.67	&	7.4	&	6.93	&	7.45	\\
	&	1 $\sEdg$ 	($\pi\rightarrow \pi^*$)	&	8.08	&	8.78	&	8.68	&	8.61	&	8.4	&	8.43	\\
   \hline
Pyridine	&	1 $\sBd$	($\pi\rightarrow \pi^*$)	&	5.06	&	5.58	&	5.46	&	5.42	&	5.31	&	5.15	\\
	&	1 $\sBj$	($n\rightarrow \pi^*$)	&	5.19	&	5.38	&	5.32	&	5.31	&	5.19	&	5.05	\\
	&	1 $\sAd$	($n\rightarrow \pi^*$)	&	5.94	&	5.73	&	5.7	&	5.66	&	5.36	&	5.5	\\
	&	2 $\sAj$	($\pi\rightarrow \pi^*$)	&	7.97	&	7.29	&	7.59	&	6.93	&	6.7	&	6.85	\\
	&	3 $\sAj$	($\pi\rightarrow \pi^*$)	&	8.47	&	8.44	&	8.55	&	8.22	&	7.86	&	7.7	\\
	&	2 $\sBd$	($\pi\rightarrow \pi^*$)	&	9.42	&	8.23	&	8.96	&	7.87	&	7.5	&	7.59	\\
   \hline   
Pyrazine	&	1 $\sBtu$	($n\rightarrow \pi^*$)	&	4.83	&	4.49	&	4.39	&	4.32	&	4.09	&	4.24	\\
	&	1 $\sAu$	($n\rightarrow \pi^*$)	&	5.99	&	5.27	&	5.2	&	5.06	&	4.67	&	5.05	\\
	&	1 $\sBdu$	($\pi\rightarrow \pi^*$)	&	4.98	&	5.43	&	5.33	&	5.27	&	5.17	&	5.02	\\
	&	1 $\sBdg$	($n\rightarrow \pi^*$)	&	5.81	&	6.03	&	6.07	&	6.01	&	5.77	&	5.74	\\
	&	1 $\sBjg$	($n\rightarrow \pi^*$)	&	7.19	&	7.13	&	7.17	&	7.02	&	6.57	&	6.75	\\
	&	1 $\sBju$	($\pi\rightarrow \pi^*$)	&	8.43	&	7.39	&	8.21	&	7.08	&	6.61	&	7.07	\\
	&	2 $\sBju$	($\pi\rightarrow \pi^*$)	&	10.29	&	8.4	&	9.92	&	8.44	&	7.53	&	8.06	\\
	&	2 $\sBdu$	($\pi\rightarrow \pi^*$)	&	9.87	&	8.36	&	9.56	&	8.34	&	7.56	&	8.05	\\
   \hline
Pyrimidine	&	1 $\sBj$	($n\rightarrow \pi^*$)	&	5.2	&	4.41	&	4.37	&	4.33	&	4.17	&	4.5	\\
	&	1 $\sAd$	($n\rightarrow \pi^*$)	&	5.82	&	4.81	&	4.83	&	4.74	&	4.48	&	4.93	\\
	&	1 $\sBd$	($\pi\rightarrow \pi^*$)	&	5.33	&	5.75	&	5.59	&	5.61	&	5.44	&	5.36	\\
	&	2 $\sAj$	($\pi\rightarrow \pi^*$)	&	7.87	&	7.52	&	7.9	&	7.45	&	7.3	&	7.06	\\
   \hline
Pyridazine	&	1 $\sBj$	($n\rightarrow \pi^*$)	&	4.25	&	3.86	&	3.89	&	3.8	&	3.69	&	3.92	\\
	&	1 $\sAd$	($\pi\rightarrow \pi^*$)	&	4.75	&	4.46	&	4.53	&	4.42	&	4.37	&	4.49	\\
	&	2 $\sAj$	($\pi\rightarrow \pi^*$)	&	5.24	&	5.61	&	5.46	&	5.47	&	5.33	&	5.22	\\
	&	2 $\sAd$	($n\rightarrow \pi^*$)	&	6.31	&	5.92	&	5.89	&	5.81	&	5.48	&	5.74	\\
   \hline
       All-E-Hexatriene	&	1 $\sBu$	($\pi\rightarrow \pi^*$)	&	7.15	&	5.71	&	6.42	&	5.66	&	5.32	&	5.58	\\
	&	2 $\sAg$	($\pi\rightarrow \pi^*$)	&	5.6	&	5.79	&	5.69	&	5.69	&	5.5	&	5.72	\\
   \hline
Tetrazine	&	1 $\sBtu$	($n\rightarrow \pi^*$)	&	3.14	&	2.78	&	2.57	&	2.47	&	2.34	&	2.53	\\
	&	1 $\sAu$	($\pi\rightarrow \pi^*$)	&	4.62	&	4.23	&	4.03	&	3.88	&	3.64	&	3.79	\\
	&	1 $\sBjg$	($n\rightarrow \pi^*$)	&	5.5	&	5.57	&	5.35	&	5.27	&	5.04	&	4.97	\\
	&	1 $\sBdu$	($\pi\rightarrow \pi^*$)	&	4.96	&	5.59	&	5.46	&	5.43	&	5.33	&	5.12	\\
	&	1 $\sBdg$	($n\rightarrow \pi^*$)	&	5.39	&	5.81	&	5.62	&	5.55	&	5.48	&	5.34	\\
	&	2 $\sAu$	($n\rightarrow \pi^*$)	&	6.24	&	5.91	&	5.62	&	5.46	&	5.04	&	5.46	\\
	&	2 $\sBdg$	($n\rightarrow \pi^*$)	&	6.94	&	7.15	&	6.9	&	6.69	&	6.19	&	6.23	\\
	&	2 $\sBjg$	($n\rightarrow \pi^*$)	&	7.21	&	7.54	&	6.89	&	6.84	&	6.64	&	6.87	\\
	&	3 $\sBjg$	($n\rightarrow \pi^*$)	&	7.69	&	7.96	&	7.53	&	7.33	&	6.79	&	7.08	\\
	&	2 $\sBtu$	($n\rightarrow \pi^*$)	&	7.54	&	7.41	&	6.96	&	6.86	&	6.59	&	6.67	\\
	&	1 $\sBju$	($\pi\rightarrow \pi^*$)	&	9.19	&	8	&	8.79	&	7.74	&	7.02	&	7.45	\\
	&	2 $\sBju$	($\pi\rightarrow \pi^*$)	&	10.06	&	8.13	&	9.72	&	8.19	&	7.11	&	7.79	\\
	&	2 $\sBdu$	($\pi\rightarrow \pi^*$)	&	8.72	&	8.95	&	8.86	&	8.68	&	8.21	&	8.51	\\
	&	2 $\sBtg$	($\pi\rightarrow \pi^*$)	&	7.92	&	9.01	&	8.79	&	8.69	&	8.59	&	8.47	
 \label{SI-tab:all-exc}
\end{longtable}
\end{center}
\bibliography{supplement}